\documentclass[conference]{IEEEtran}
\IEEEoverridecommandlockouts
\usepackage[backend=bibtex, style=ieee]{biblatex} 
\usepackage{amsmath,amssymb,amsfonts}
\usepackage{algorithmic}
\usepackage{graphicx}
\usepackage{textcomp}
\usepackage{subcaption}
\usepackage{xcolor}
\usepackage{url}
\usepackage{hyperref}
\def\BibTeX{{\rm B\kern-.05em{\sc i\kern-.025em b}\kern-.08em
    T\kern-.1667em\lower.7ex\hbox{E}\kern-.125emX}}
\begin{document}

\title{Manim for STEM Education: Visualizing Complex Problems Through Animation\\
}

\author{\IEEEauthorblockN{Christina Zhang}
\IEEEauthorblockA{\textit{Departments of Computer Science and Mathematics} \\
\textit{Purdue University}\\
305 N University St, West Lafayette, IN 47907, USA \\
zhan4727@purdue.edu}
}

\maketitle

\begin{abstract}
Many STEM concepts pose significant learning challenges to students due to their inherent complexity and abstract nature. Visualizing complex problems through animations can significantly enhance learning outcomes. However, the creation of animations can be time-consuming and inconvenient. Hence, many educators illustrate complex concepts by hand on a board or a digital device. Although static graphics are helpful for understanding, they are less effective than animations. The free and open-source Python package Manim enables educators to create visually compelling animations easily. Python’s straightforward syntax, combined with Manim’s comprehensive set of built-in classes and methods, greatly simplifies implementation. This article presents a series of examples that demonstrate how Manim can be used to create animated video lessons for a variety of topics in computer science and mathematics. In addition, it analyzes viewer feedback collected across multiple social media platforms to evaluate the effectiveness and accessibility of these visualizations. The article further explores broader potentials of the Manim Python library by showcasing demonstrations that extend its applications to subject areas beyond computer science and mathematics.
\end{abstract}


\section{Introduction}
Many STEM topics are difficult to learn. Visualization often helps significantly to understand these concepts \cite{mayer2002animation}. Previous research shows that animations are more effective than static graphics for enhancing learning outcome \cite{berney2016does}. Animations offer students graphical representations that illustrate complex problems and also allow students to engage in the learning process. However, creating animations can be extremely time-consuming and inconvenient for educators. Hence, educators mostly rely on static graphics when teaching, hand-drawing visualizations for complex problems on boards or digital devices. In addition, some educators try to find animations available online to show their students, but there may not be well-made animations posted online for specific examples that the educator finds necessary to explain. 

A free and open-source Python package called Manim allows educators to easily create highly polished and visually compelling animations. The package was developed by Grant Sanderson, the creator of the 3Blue1Brown YouTube channel \cite{The_Manim_Community_Developers_Manim_Mathematical_2025}. It contains many built-in functions that naturally create animations of objects and embeds \LaTeX\  for writing mathematical equations, allowing users to create clear, professional-looking animations with minimal effort. Installation of the Manim package is straightforward. It is sufficient to write code using a basic text editor such as Windows Notepad. Compared to programming languages such as C and Java, Python is much easier to learn. Creating Manim animations---aside from extremely advanced use cases---often only requires knowledge of a small subset of Python syntax. This accessibility opens the tool to educators, students, and STEM professionals with little programming background.

Manim has proven to be effective in educational contexts. A 2024 study by Milorad Marković and Ivan Kaštelan demonstrated its success in visualizing data structures and algorithms in an undergraduate course, with most of the participants reporting that the animations significantly improved their conceptual understanding of the topics \cite{markovic2024demonstrating}. Another 2024 study examined the use of the tool for visualizing parsing algorithms, stating that the visualizations make it easier to understand the steps of parsing \cite{akhilesh2024visual}. A 2023 pilot study also showed that Manim animations helped electrical and electronic engineering students understand abstract concepts in a Signals and Systems course \cite{so2023pilot}. Furthermore, the preparation of animation templates allows the autogeneration of explanatory math animations when only a specific math question is given as input from the user \cite{yang2022autogeneration}. Previous research shows that the integration of a step-by-step solver into the autogeneration system can help generate more types of explanatory math animations and can significantly reduce the time for animation template design \cite{huang2023efficient}. However, many educators may prefer to develop their own explanations of the problems for their students. Hence, writing their own Python code is necessary.

In this paper, I report my experience using the community edition of Manim to create animated video lessons for various topics in computer science and mathematics. The community edition of Manim was forked from Grant Sanderson's GitHub repository 3b1b/manim and developed by the community. This version is recommended for general use because of active community maintenance, feature improvements, and better documentation \cite{The_Manim_Community_Developers_Manim_Mathematical_2025}. Each video lesson I create focuses on a specific problem or topic. This paper describes the detailed process of making these videos, analyzes feedback from viewers on various social media platforms, and demonstrates the potential of using Manim to teach STEM subjects other than computer science and mathematics.

\section{Design and Implementation}

\subsection{How to Use Manim}



The community edition of Manim can be installed on a personal computer by following these steps:
\begin{enumerate}
    \item Install the latest version of Python.
    \item Install the dependencies FFmpeg (for video rendering) and \LaTeX \ (for mathematical equations).
    \item Install Manim by running \verb|pip install manim| in the terminal.
\end{enumerate}
After installation, educators can create animations by writing Python code using an IDE such as Visual Studio Code or a simple text editor such as Windows Notepad. The extension \verb|.py| specifies a Python file. The top of the file should include the line \verb|from manim import *|. The following is a simple example.
\begin{verbatim}
# example.py
from manim import *
class Example(Scene):
    def construct(self):
        text = Text("Hello World")
        self.play(Write(text))
        self.wait(2)
        self.play(FadeOut(text))
\end{verbatim}
The command \verb|manim -pql example.py Example| renders the \verb|Example| scene from \verb|example.py| in low quality (480p15) and generates an MP4 file. For higher quality videos, the flags can be set to \verb|-pqm| (720p30), \verb|-pqh| (1080p60), or \verb|-pqk| (2160p60). The message ``Hello World" is displayed on the screen using a handwriting-style animation. The video pauses for 2 seconds due to the line \verb|self.wait(2)|. Then, the text fades off the screen. Figure 1 shows screenshots of the terminal output and the rendered video. 

\begin{figure}[htbp!]
    \begin{subfigure}[b]{0.233\textwidth}
        \includegraphics[width=\linewidth]{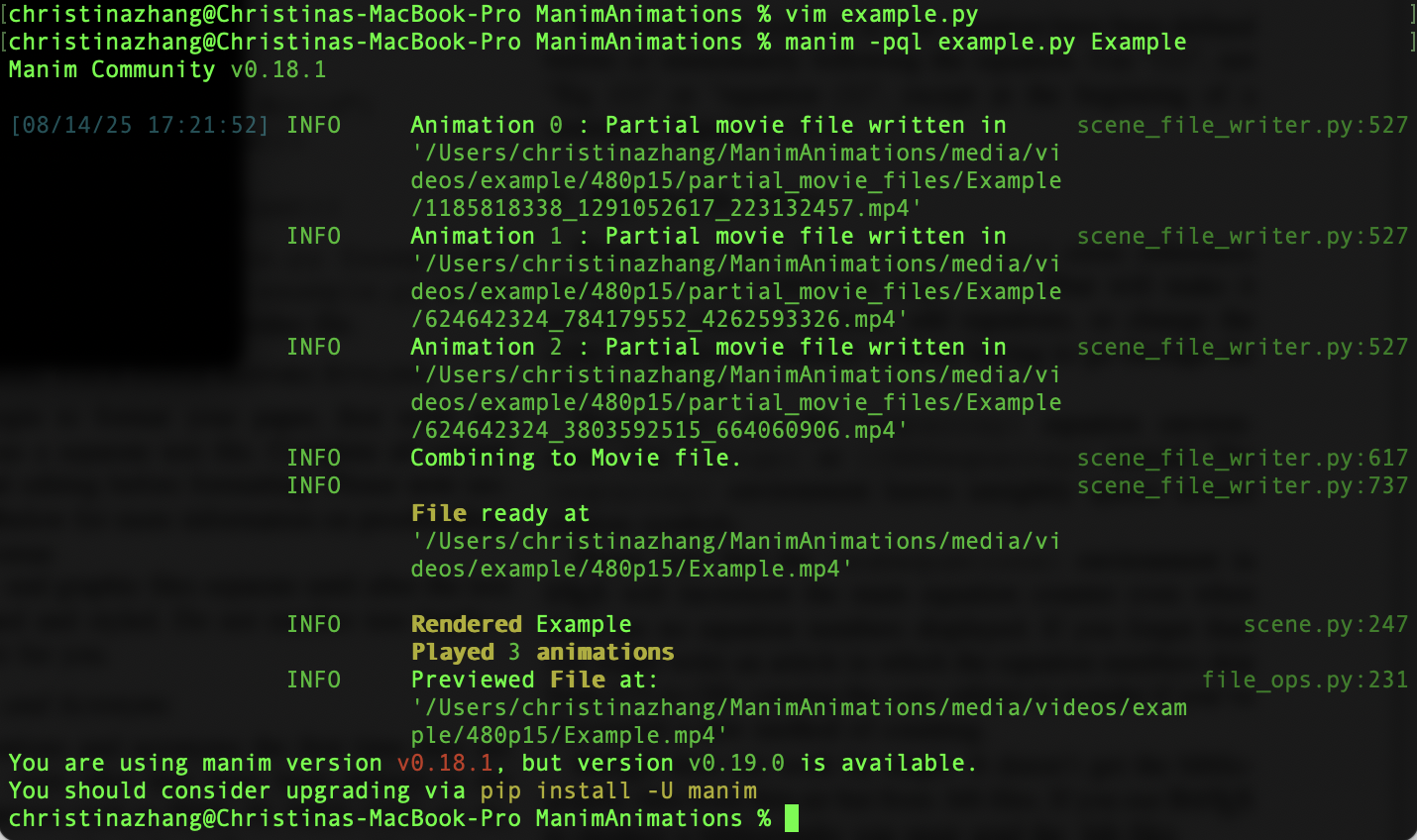}
        \caption{Terminal Output}
        \label{fig:sub1}
    \end{subfigure}
    \begin{subfigure}[b]{0.247\textwidth}
        \includegraphics[width=\linewidth]{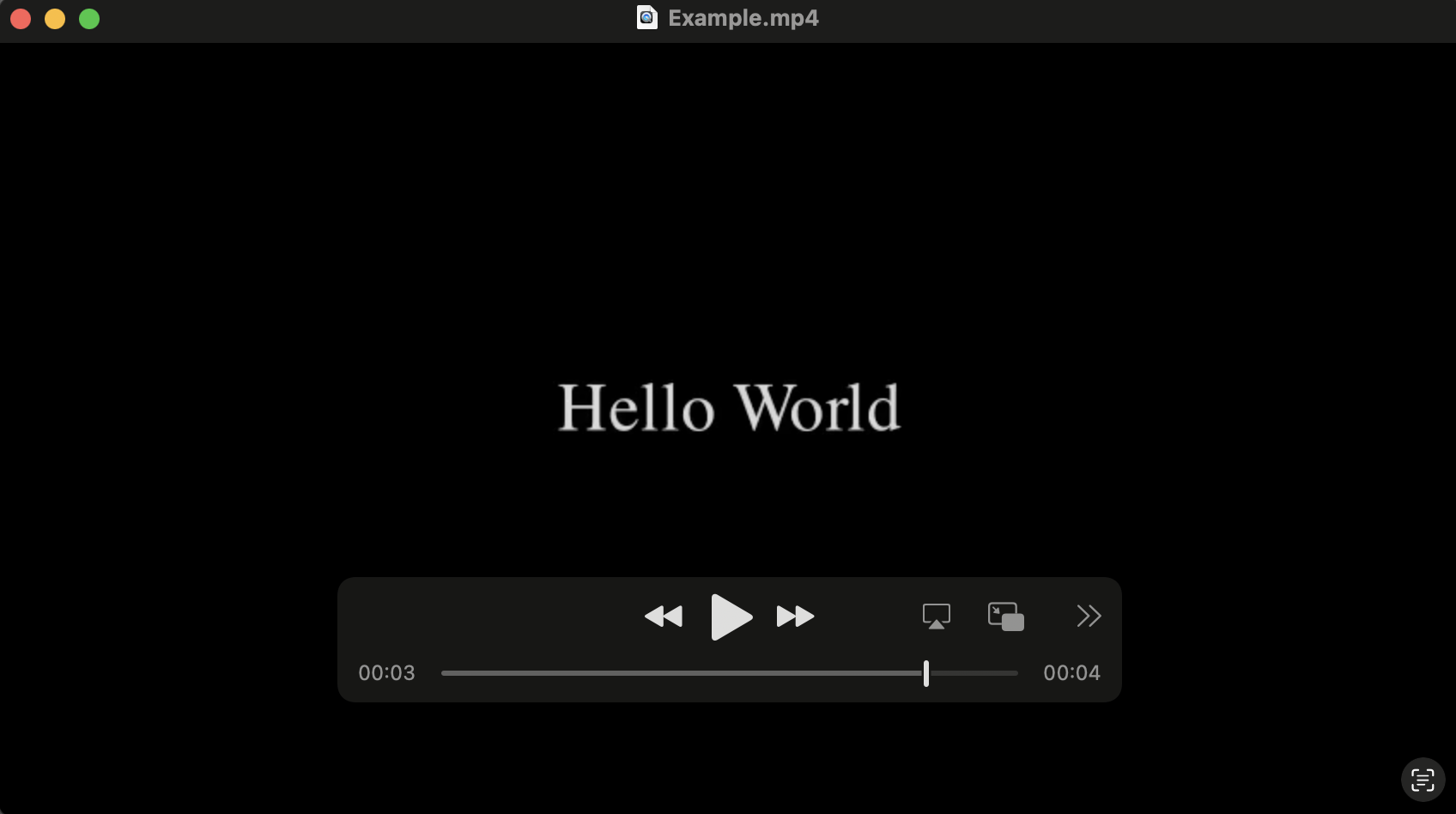}
        \caption{Rendered Video}
        \label{fig:sub2}
    \end{subfigure}
    \caption{Simple Manim animation example.}
    \label{fig:main}
\end{figure}

In Manim, any animation must be passed to the function \verb|self.play()| to be rendered on the screen \cite{The_Manim_Community_Developers_Manim_Mathematical_2025}. Multiple animations can be played at once when passed as arguments to this method. Manim contains many built-in classes such as \verb|Circle| and \verb|Rectangle|. It can display images and SVG files on screen. Every rendered Manim video corresponds to a scene class. Each scene class must implement the \verb|construct()| method, where all animations and objects for that scene are defined. There are many types of scene classes, but \verb|Scene| is sufficient for creating the majority of animated video lessons. An educator can change this to \verb|ThreeDScene| if 3-dimensional objects are required, or to \verb|MovingCameraScene| if a camera that can pan and zoom is needed.

\subsection{Creating Animated Video Lessons}

The first step in developing an animated video lesson is to precisely identify the concept or problem to be explained. This could be a sorting algorithm, a chemical reaction, a physics equation, or a proof of a theorem. Once the problem is clearly identified, the next step is to begin developing the explanations and creating the visualizations. It is important to implement the animations when planning each part of the explanation because this allows the educator to actively participate in the learning process, mirroring the student experience. A narration script is not required at this stage. It is clear through the animations what is being explained. 

After the animations have been rendered into a full video, the next step involves adding narrations. This step can be skipped if the educator plans to provide the explanation when showing the animation to the students. However, adding narrations eliminates the need for repeated instructor explanations during video playback. Unfortunately, recording narrations is very time-consuming and repetitive. A single misspoken word requires restarting the recording from the beginning. Using a text-to-speech AI model solves this problem. MiniMax AI's text-to-speech AI model generates natural-sounding voices from the user's input text for free \cite{minimax2025}. This is the model used for my videos. There are many other models to choose from. 


Figure 2 shows the process of explaining the heapsort algorithm in an animated video lesson. The steps involve explaining the heapsort pseudocode in detail, explaining the heapify pseudocode in detail, and animating the process of using heapsort to sort an array of integers. Specific lines of pseudocode are highlighted in the animation when they are explained. 

\begin{figure}[htbp!]
    \centering

    \begin{subfigure}[b]{0.235\textwidth}
        \includegraphics[width=\linewidth]{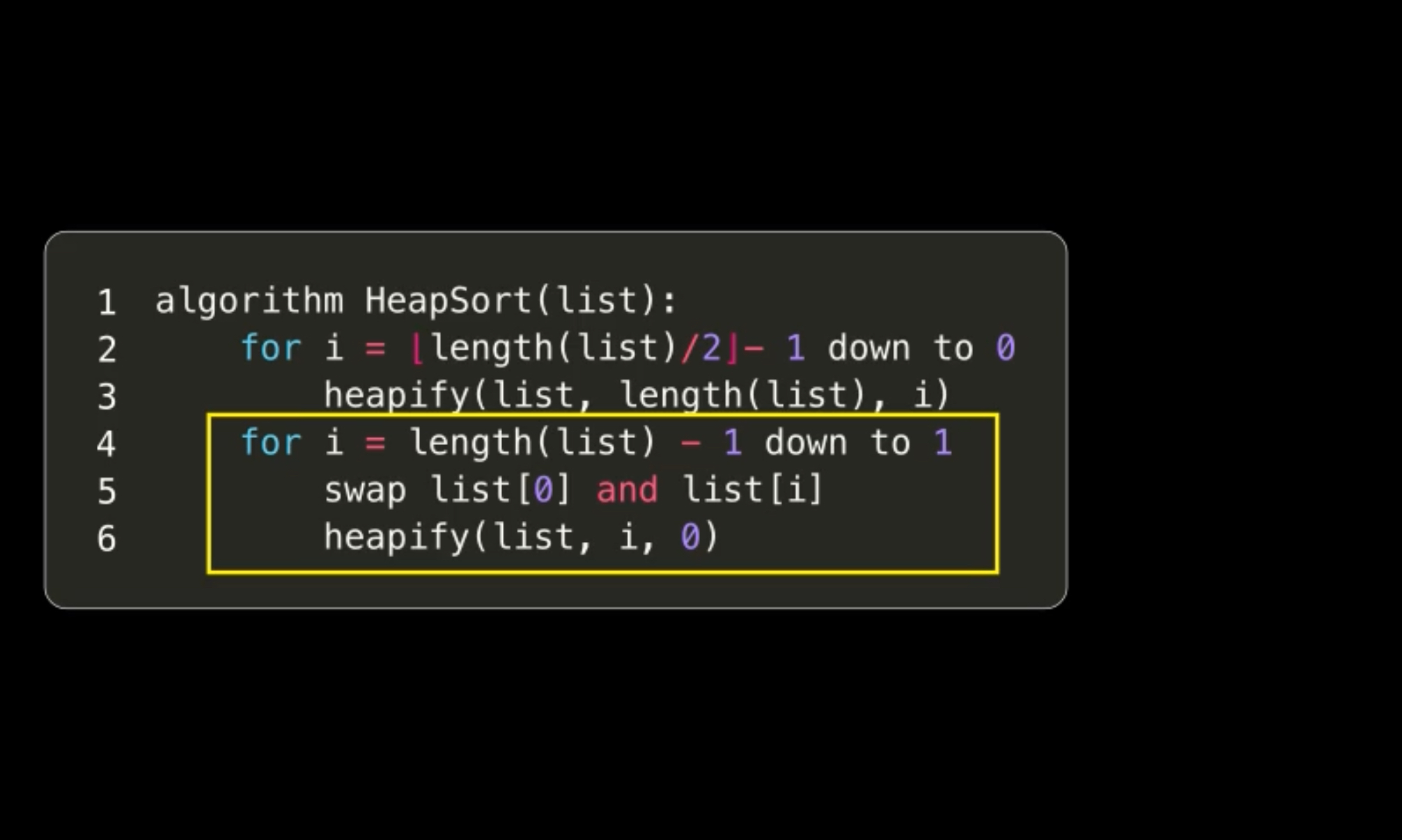}
        \caption{Heapsort Pseudocode}
        \label{fig:a}
    \end{subfigure}
    \hfill
    \begin{subfigure}[b]{0.235\textwidth}
        \includegraphics[width=\linewidth]{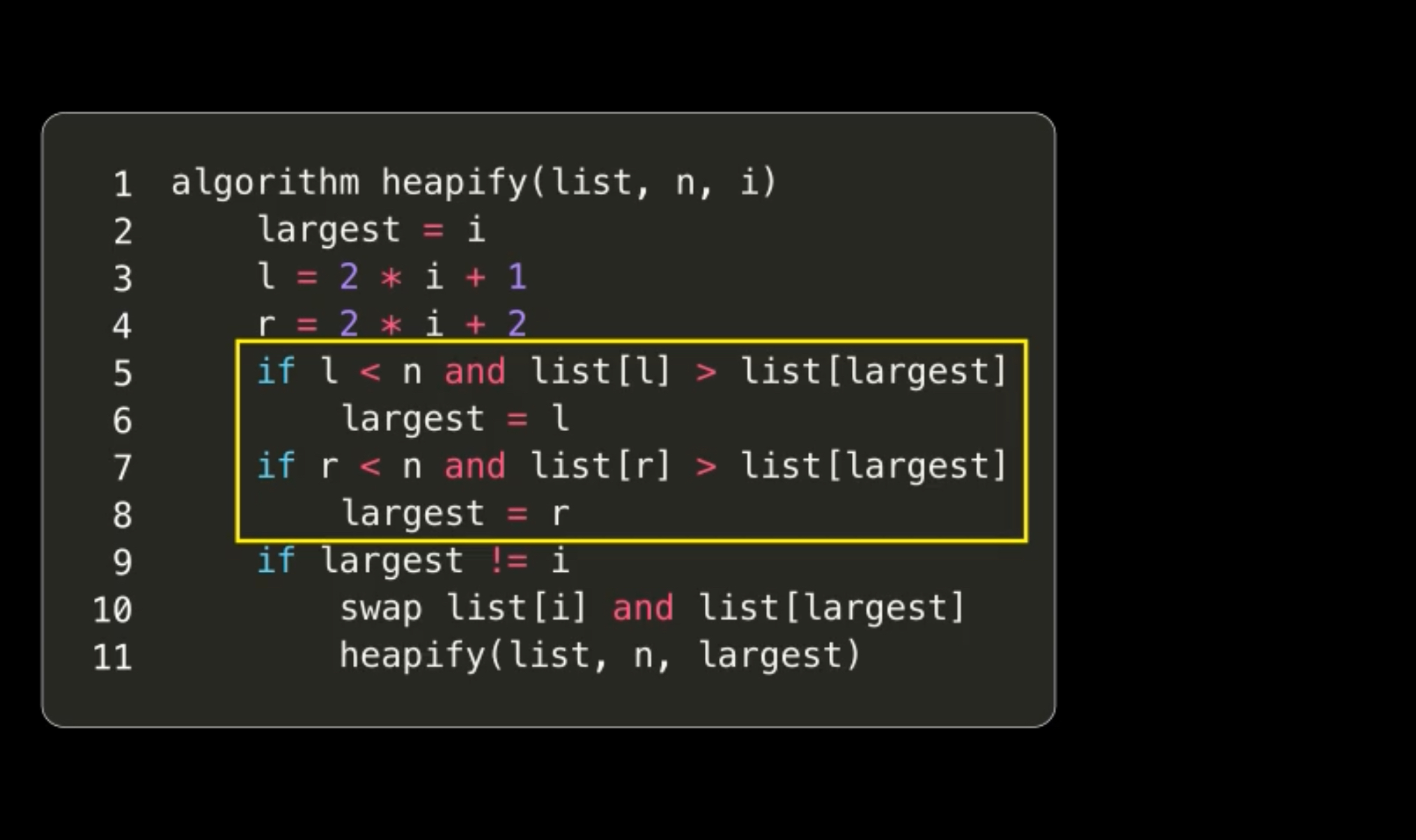}
        \caption{Heapify Pseudocode}
        \label{fig:b}
    \end{subfigure}

    \vskip\baselineskip 

    \begin{subfigure}[b]{0.24\textwidth}
        \includegraphics[width=\linewidth]{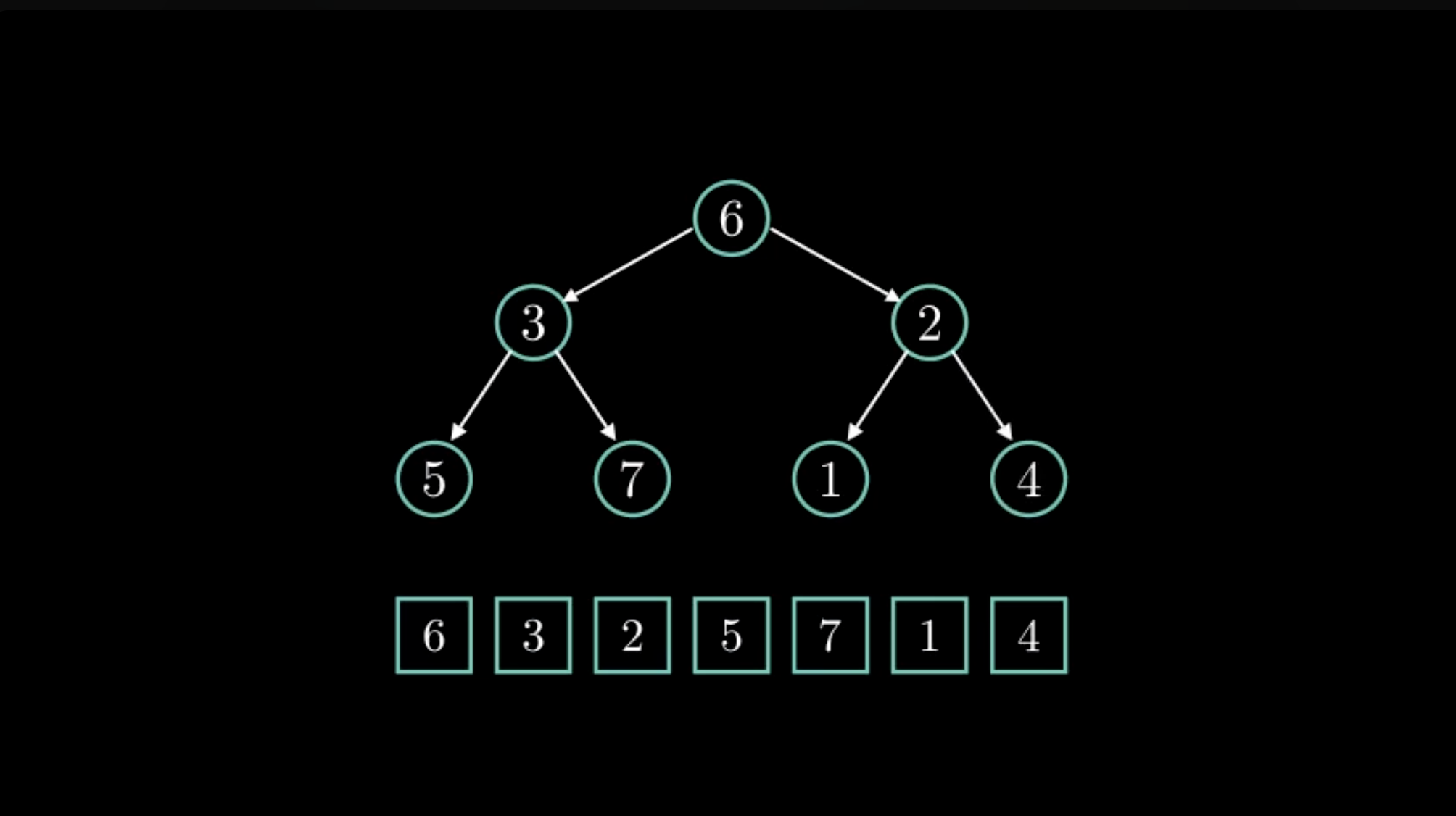}
        \caption{Initial state of Array/Heap}
        \label{fig:c}
    \end{subfigure}
    \hfill
    \begin{subfigure}[b]{0.225\textwidth}
        \includegraphics[width=\linewidth]{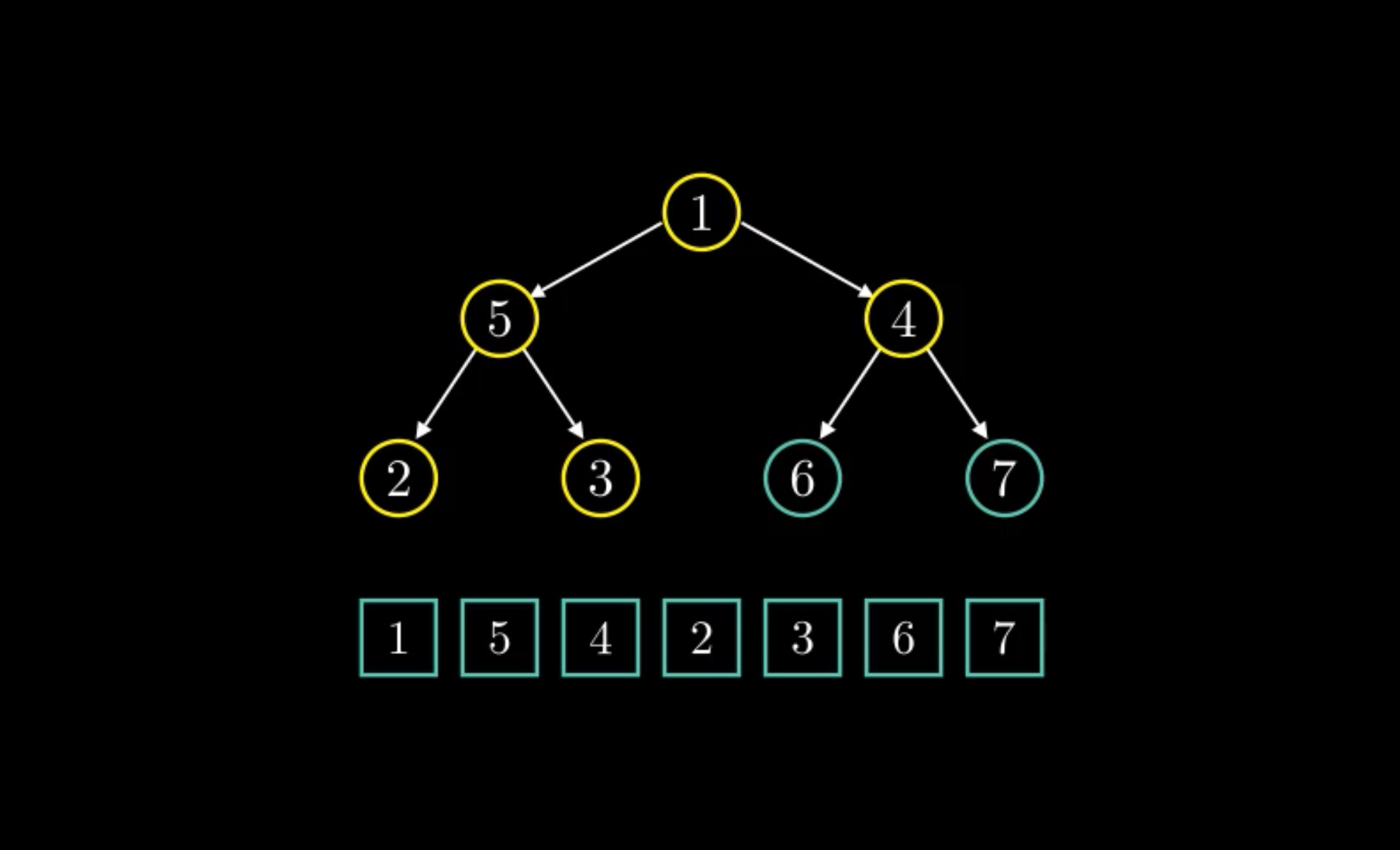}
        \caption{Sorting Process}
        \label{fig:d}
    \end{subfigure}

    \caption{Animation explaining heapsort.}
    \label{fig:grid}
\end{figure}

Heapsort sorts an array in non-decreasing order by first viewing the entries as nodes of a complete binary tree and building it into a max-heap \cite{williams1964algorithm}. In a max-heap, the maximum value is stored at the root (the topmost node). Hence, swapping the root node with the last node places the maximum value at the end of the array. The remaining nodes are built into another max-heap so that the second largest value is at the root. This process is repeated until the entire array is sorted. In the heapsort video, two animations are played at once for each swap: the swapping of nodes in the tree, and the swapping of entries in the array. Showing both animations allows students to visualize the swap in the tree data structure and the corresponding swap in the array to be sorted. 

Figure 3 displays selected frames from an animation explaining the Gaussian Integral $\int_{-\infty}^{\infty}e^{-x^2}\ dx$. The value of this integral is the area under the curve $f(x)=e^{-x^2}$ over the real line. The animation sequentially: (1) plots the function, (2) highlights the area under the curve, (3) displays each derivation step, and (4) concludes by boxing the final result $\int_{-\infty}^{\infty}e^{-x^2}\ dx=\sqrt{\pi}$.

\begin{figure}[htbp!]
    \begin{subfigure}[b]{0.15\textwidth}
        \includegraphics[width=\linewidth]{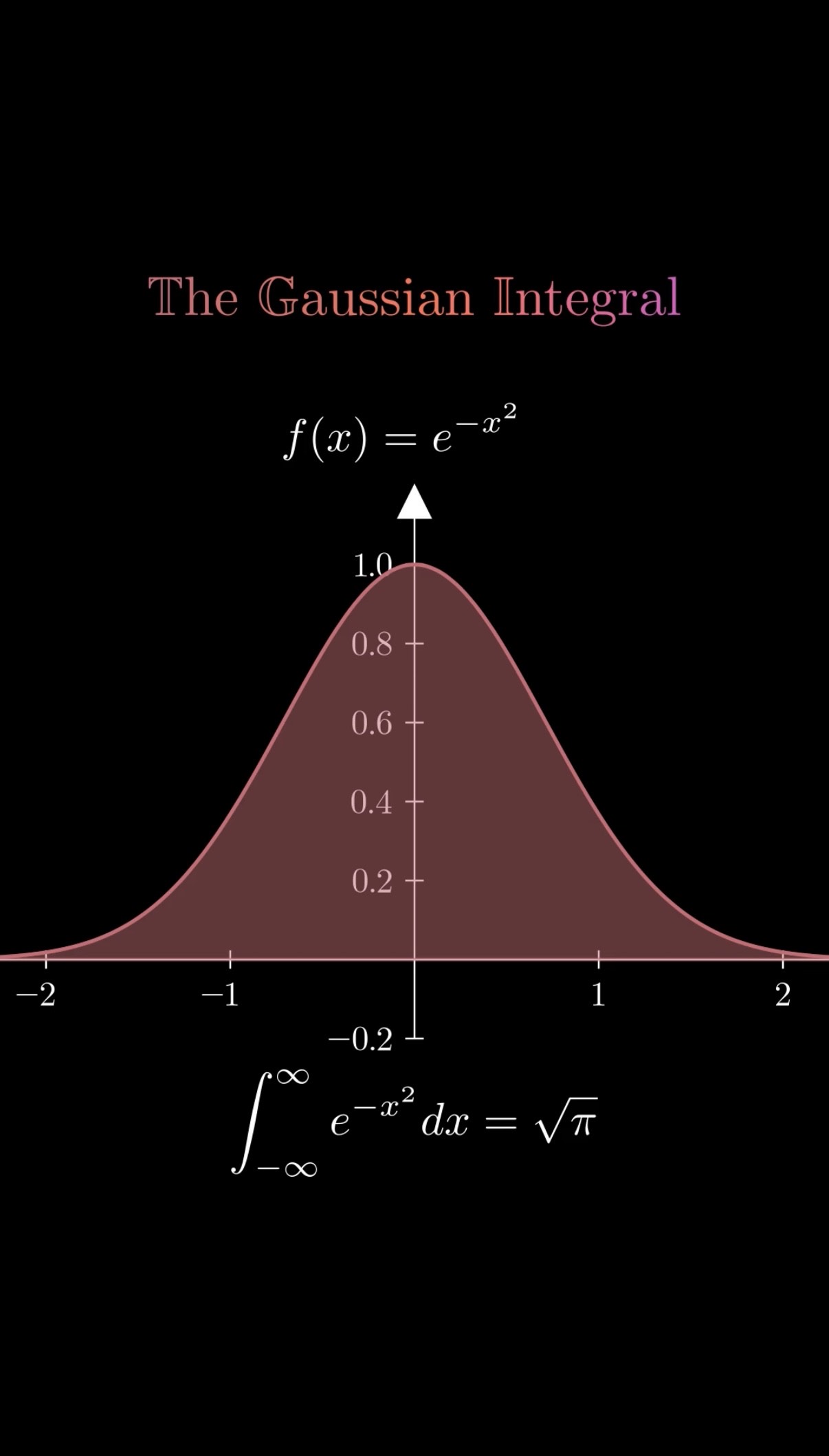}
        \caption{Visualization}
        \label{fig:sub1}
    \end{subfigure}
    \begin{subfigure}[b]{0.15\textwidth}
        \includegraphics[width=\linewidth]{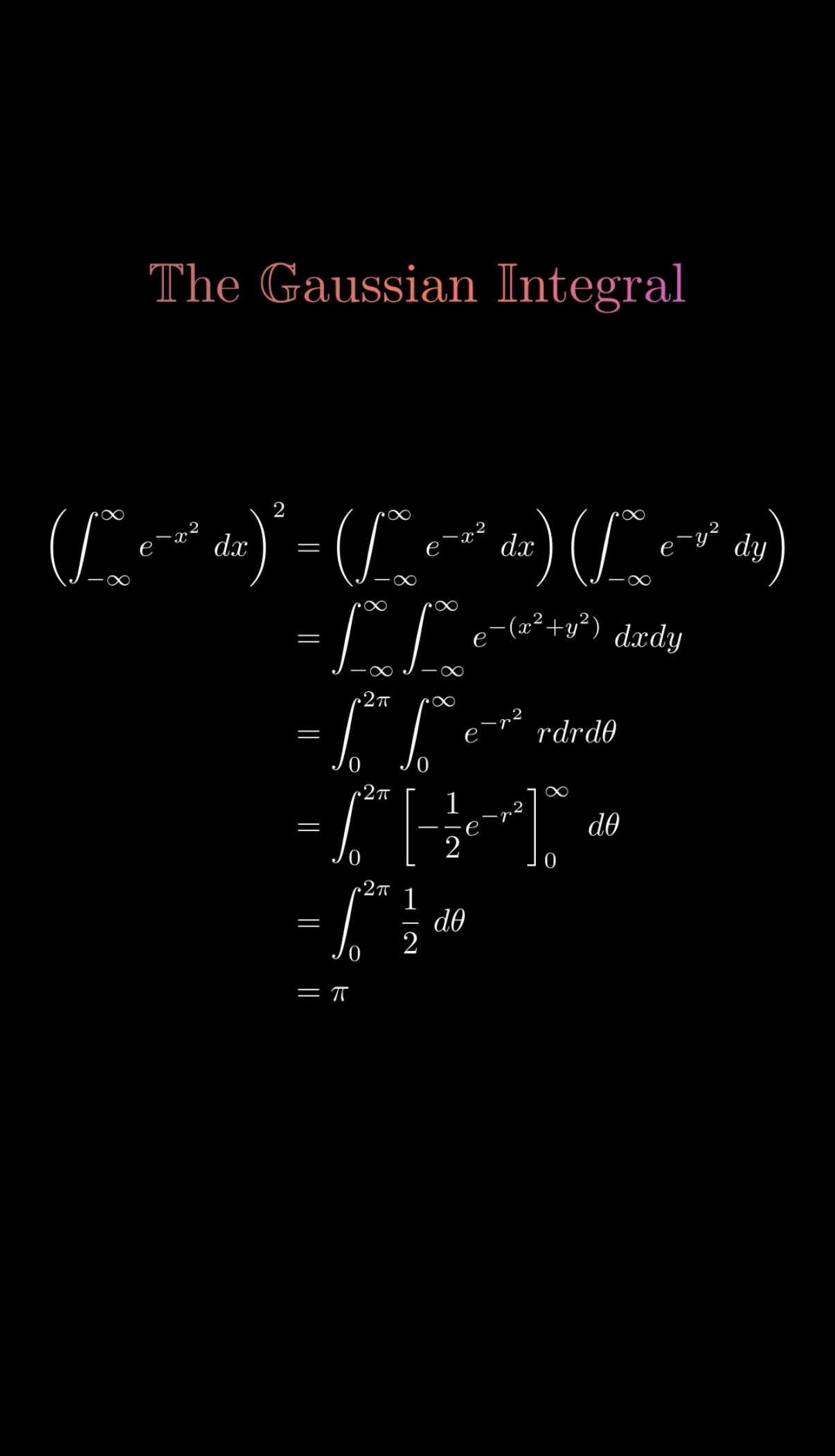}
        \caption{Derivation}
        \label{fig:sub2}
    \end{subfigure}
    \begin{subfigure}[b]{0.15\textwidth}
        \includegraphics[width=\linewidth]{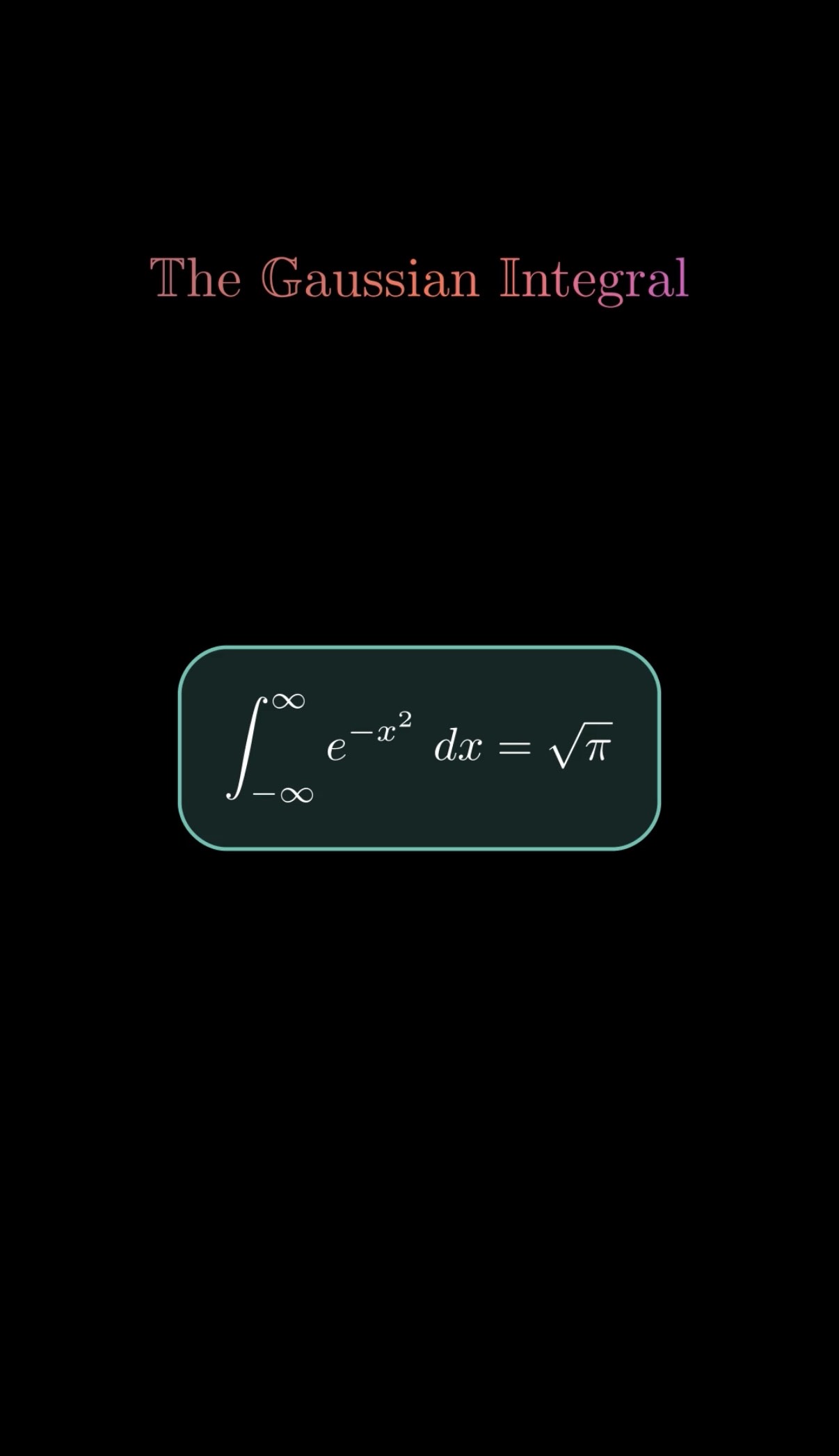}
        \caption{Boxed Answer}
        \label{fig:sub2}
    \end{subfigure}
    \caption{Animation explaining the Gaussian Integral.}
    \label{fig:main}
\end{figure}

In some cases, it is more appropriate to use a white background. The background color is set to black by default. It can be overwritten in the Python code to other colors. See Figure 4. 

\begin{figure}[htbp!]
    \centering

    \begin{subfigure}[b]{0.23\textwidth}
        \includegraphics[width=\linewidth]{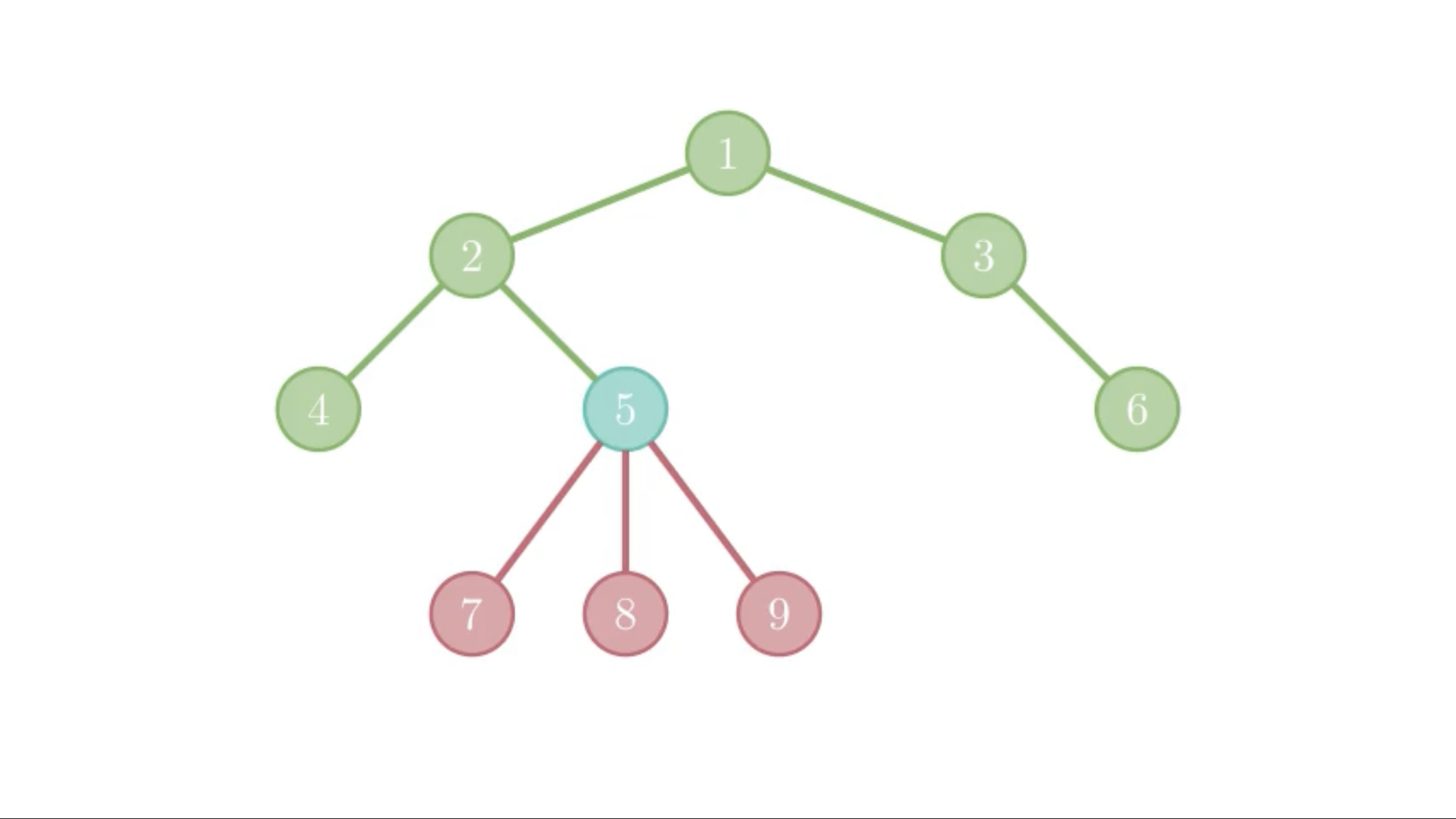}
        \caption{Defining a Tree}
        \label{fig:a}
    \end{subfigure}
    \hfill
    \begin{subfigure}[b]{0.23\textwidth}
        \includegraphics[width=\linewidth]{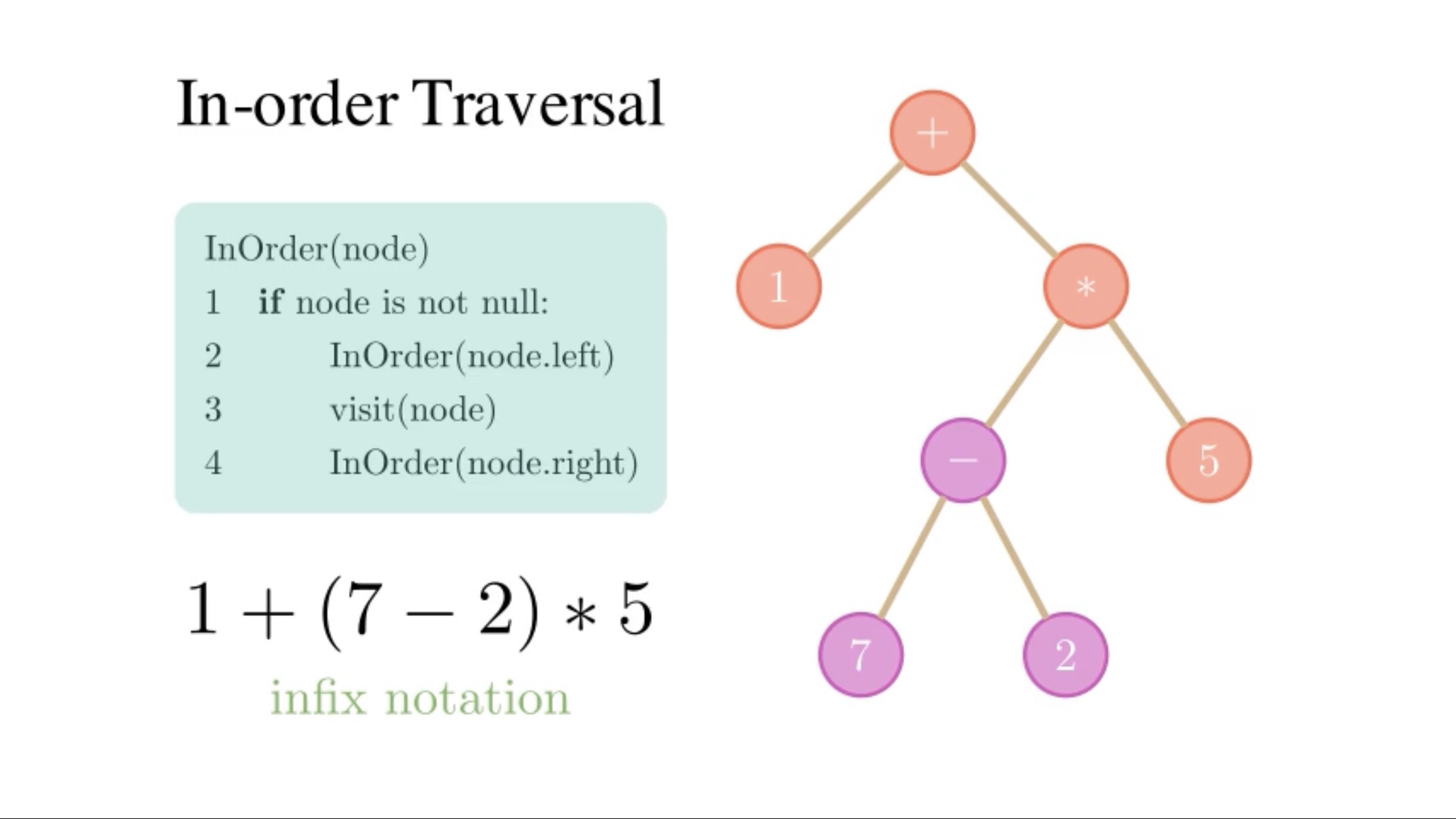}
        \caption{In-order Traversal}
        \label{fig:b}
    \end{subfigure}

    \vskip\baselineskip 

    \begin{subfigure}[b]{0.23\textwidth}
        \includegraphics[width=\linewidth]{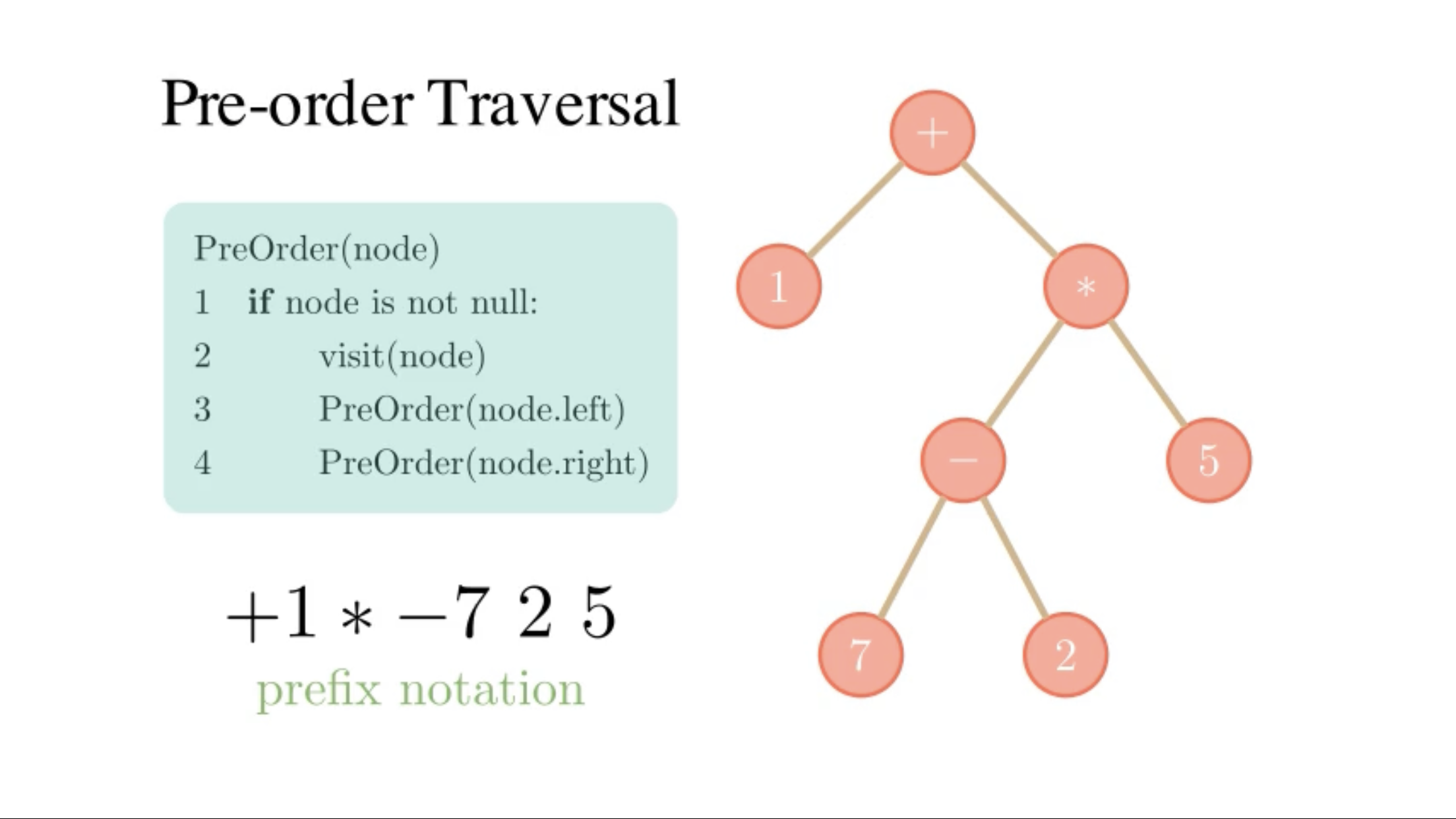}
        \caption{Pre-order Traversal}
        \label{fig:c}
    \end{subfigure}
    \hfill
    \begin{subfigure}[b]{0.23\textwidth}
        \includegraphics[width=\linewidth]{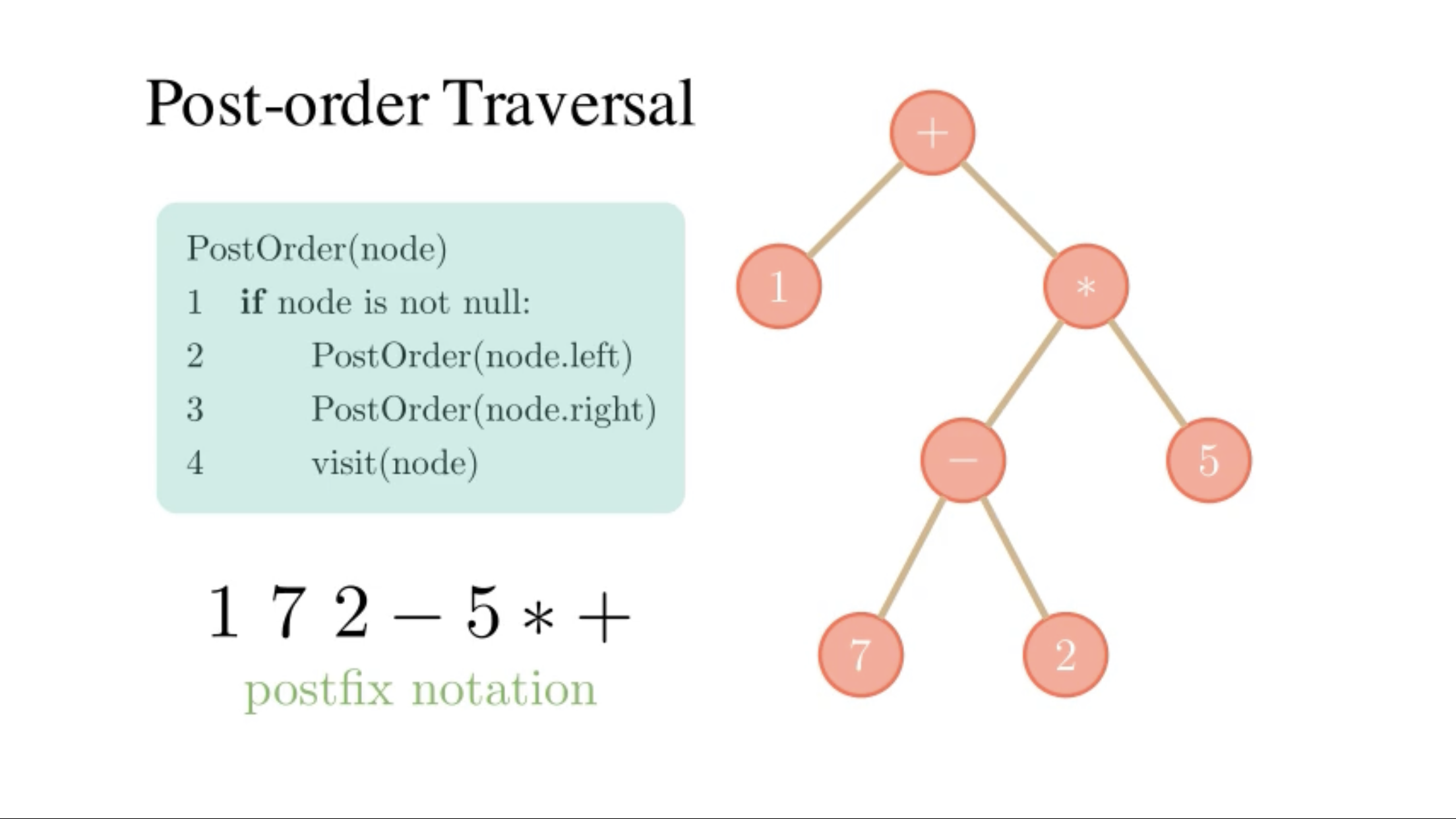}
        \caption{Post-order Traversal}
        \label{fig:d}
    \end{subfigure}

    \caption{Animation explaining the tree data structure.}
    \label{fig:grid}
\end{figure}

Figure 5 shows selected frames from two videos that include the visualization of a determinant and an ellipsoid, respectively. The built-in methods and classes make the construction of these scenes straightforward.

\begin{figure}[htbp!]
    \centering
    \begin{subfigure}[b]{0.24\textwidth}
        \includegraphics[width=\linewidth]{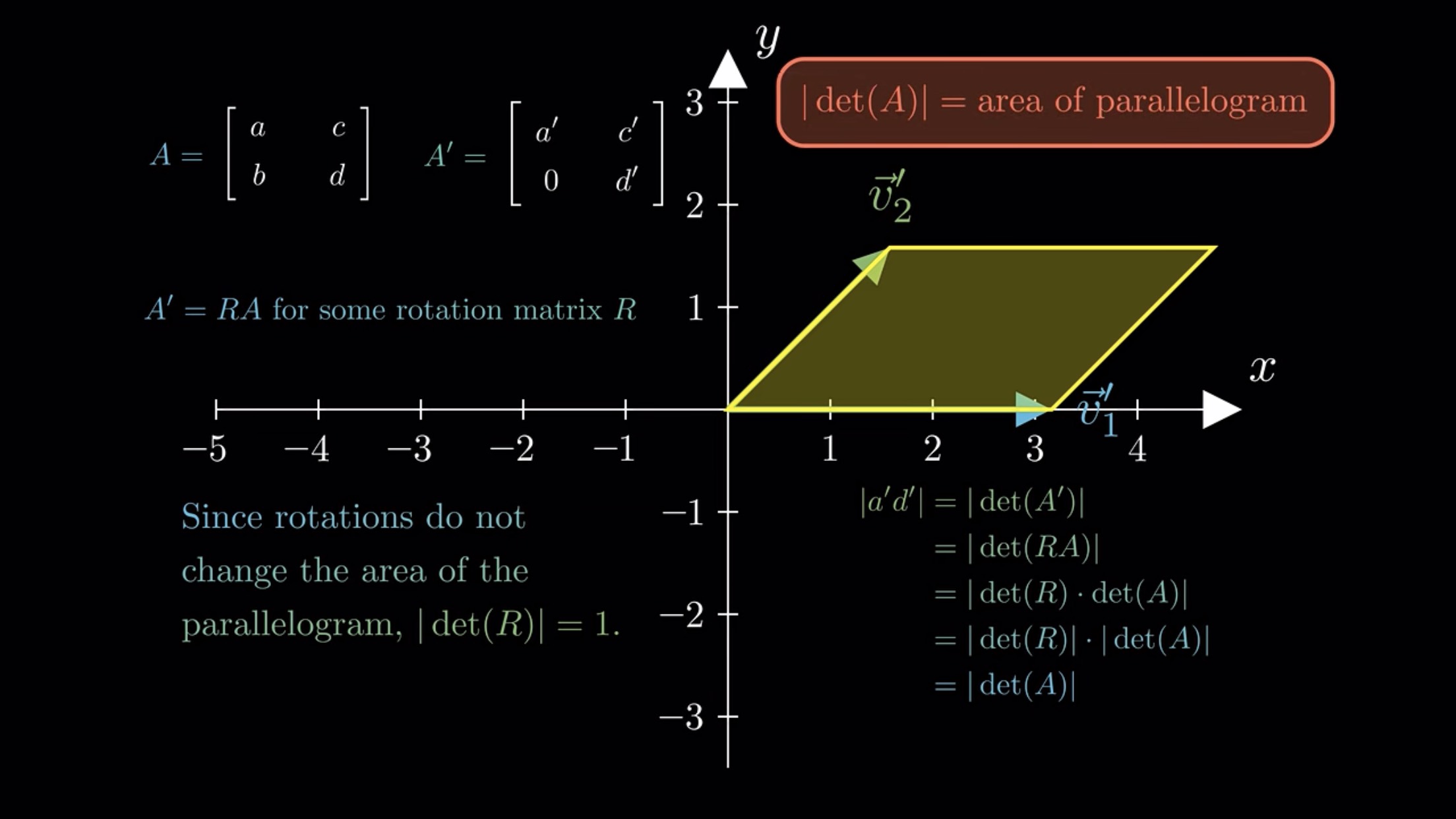}
        \caption{Determinant Visualization}
        \label{fig:a}
    \end{subfigure}
    \hfill
    \begin{subfigure}[b]{0.24\textwidth}
        \includegraphics[width=\linewidth]{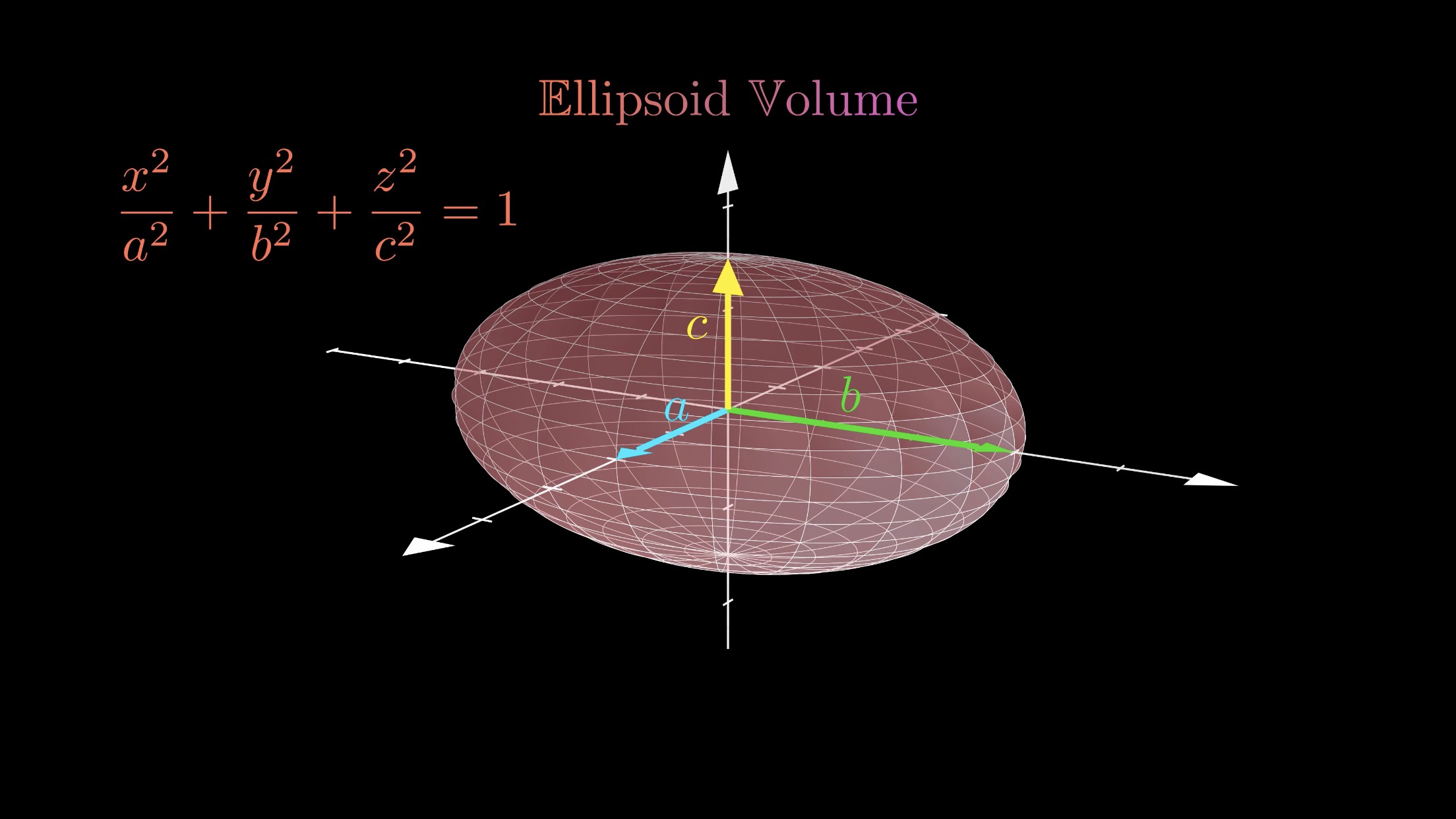}
        \caption{Ellipsoid Visualization}
        \label{fig:b}
    \end{subfigure}
    
    \caption{2D determinant and 3D parametric surface.}
    \label{fig:grid}
\end{figure}

\section{Feedback From Viewers}

The animated video lessons were shared on YouTube, TikTok, RedNote, and BiliBili. The viewers provided feedback through likes and comments. This section analyzes the feedback and suggests improvements for future videos.

Figure 6 shows key frames from a video that derives the closed-form formula for $\sum_{k=0}^{N}\cos k\theta$. This was the first video to reach over a thousand views on my YouTube channel. The channel had less than 60 subscribers when the video passed a thousand views in 3 days. Viewers of the video left many encouraging comments praising the elegance shown on screen. The video’s success stemmed from its engaging structure: (1) the problem was introduced as a question, (2) two important formulas used in the proof were listed upfront before the scene transitions to the detailed derivation steps, and (3) the derived formula is boxed at the end for emphasis. The use of instrumental background music further enhanced engagement, a practice worth retaining.

\begin{figure}[htbp!]
    \centering

    \begin{subfigure}[b]{0.225\textwidth}
        \includegraphics[width=\linewidth]{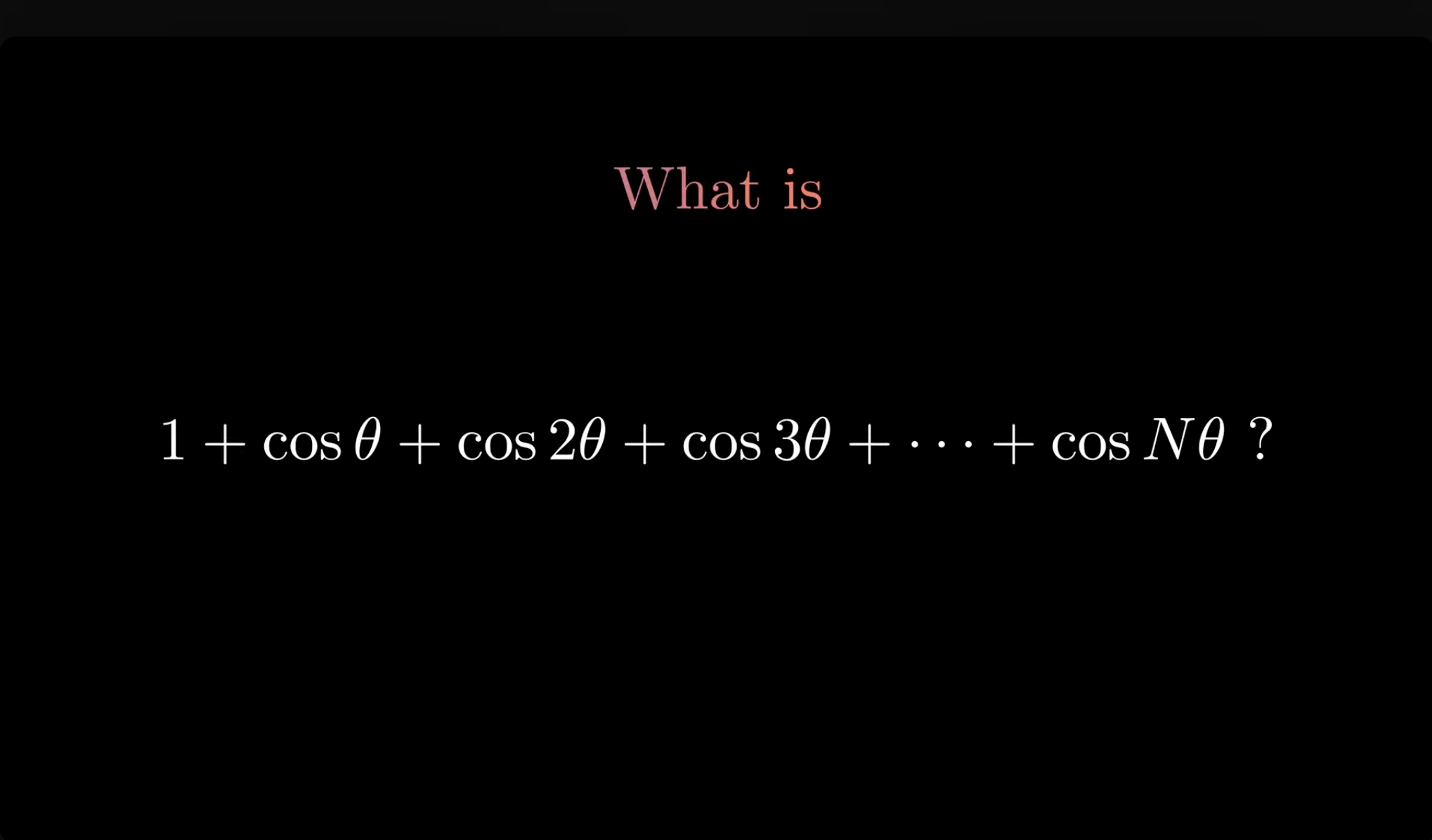}
        \caption{Asking the Question}
        \label{fig:a}
    \end{subfigure}
    \hfill
    \begin{subfigure}[b]{0.23\textwidth}
        \includegraphics[width=\linewidth]{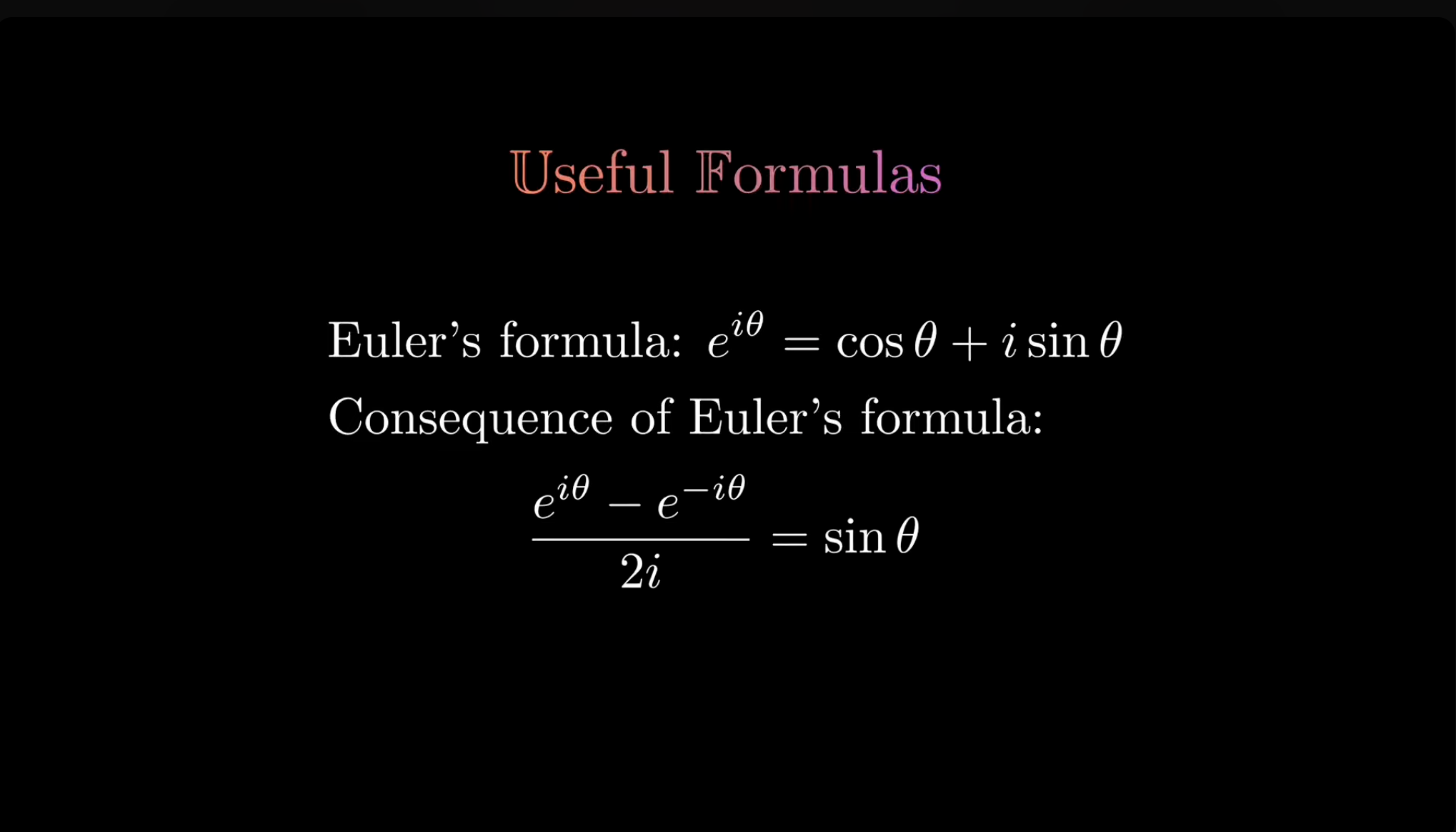}
        \caption{List of Useful Formulas}
        \label{fig:b}
    \end{subfigure}

    \vskip\baselineskip 

    \begin{subfigure}[b]{0.225\textwidth}
        \includegraphics[width=\linewidth]{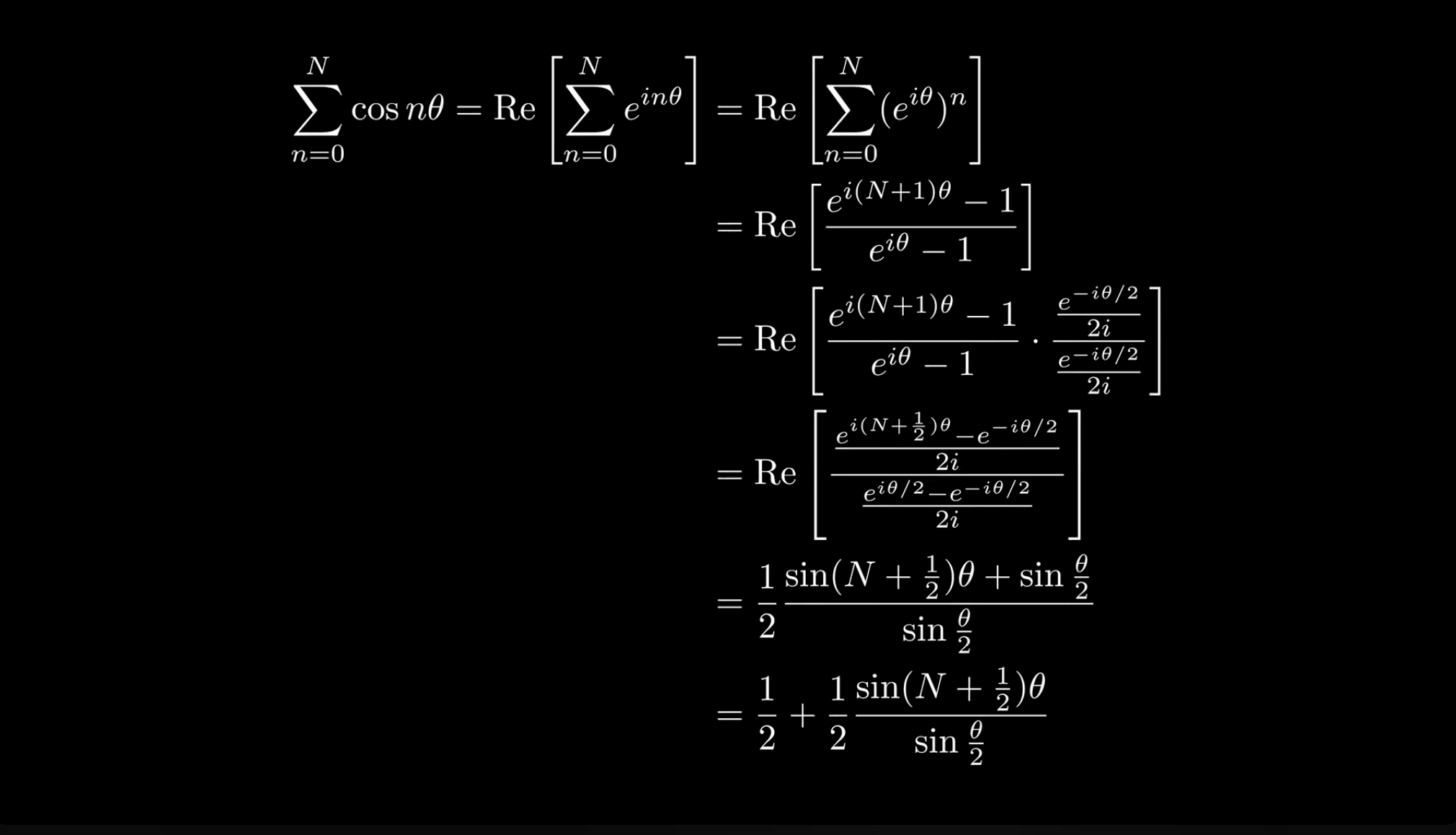}
        \caption{Derivation Steps}
        \label{fig:c}
    \end{subfigure}
    \hfill
    \begin{subfigure}[b]{0.23\textwidth}
        \includegraphics[width=\linewidth]{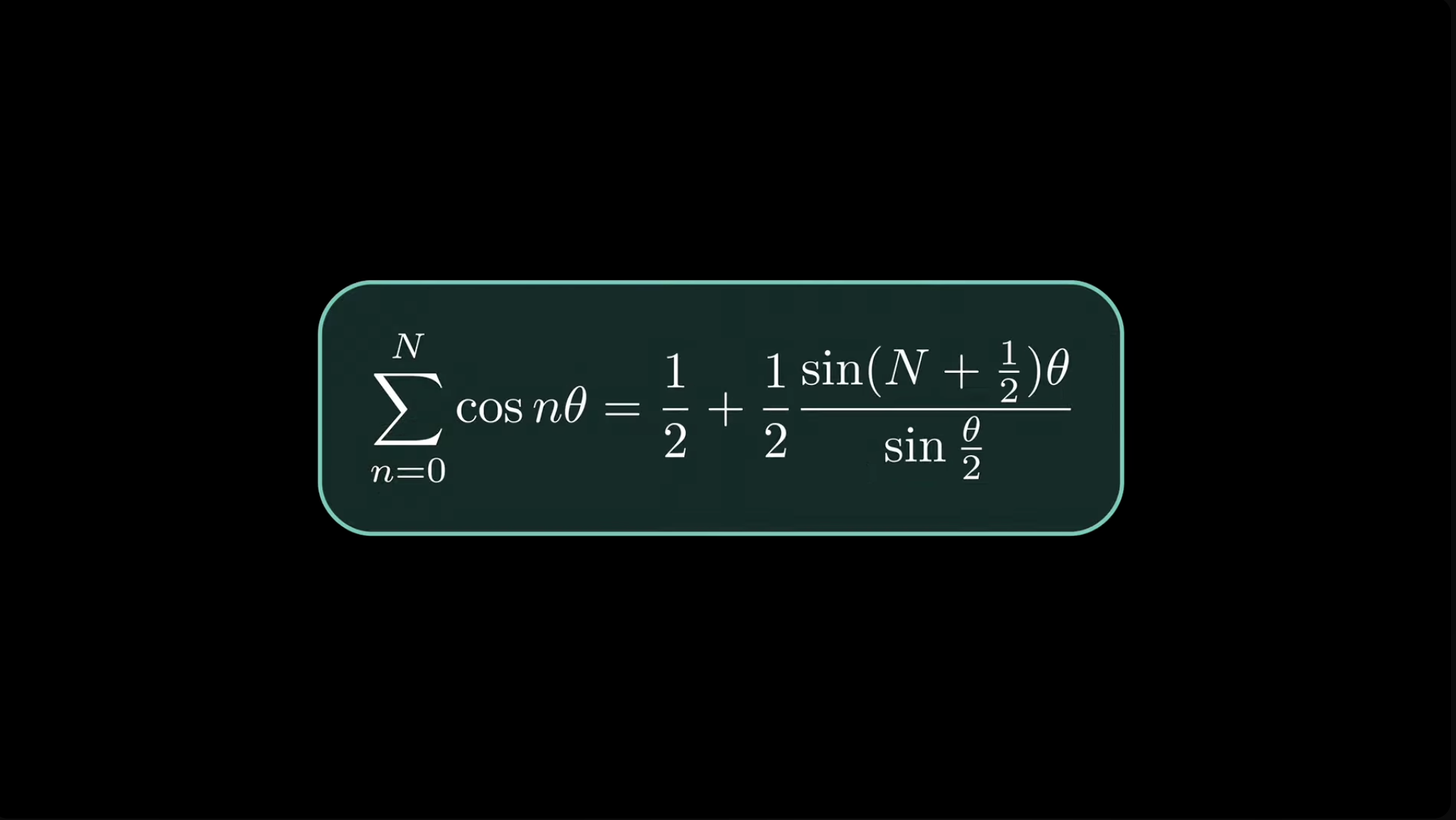}
        \caption{Boxed Formula}
        \label{fig:d}
    \end{subfigure}

    \caption{Derivation of the formula for $\sum_{k=0}^{N}\cos k\theta$.}
    \label{fig:grid}
\end{figure}

A few videos on data structures and algorithms surpassed 100 likes on TikTok within days. Many viewers also chose to save the videos. Figure 4 contains frames from a video that has more than 200 likes on TikTok. These videos are only 1 to 3 minutes long but cover the topic in a comprehensive way. The combination of animations, narrations, and instrumental background music makes the topic more interesting and easier to understand. 

Some viewers suggested improvements in the comments section. For instance, one TikTok viewer suggested using a color other than teal for the visualization in Figure 7 because it was difficult to read the labels on the nodes. Another added the suggestion of using a bolder font. These minor improvements can enhance the overall viewing experience and should be applied to future videos. 

\begin{figure}[htbp!]
\centerline{\includegraphics[width=8cm]{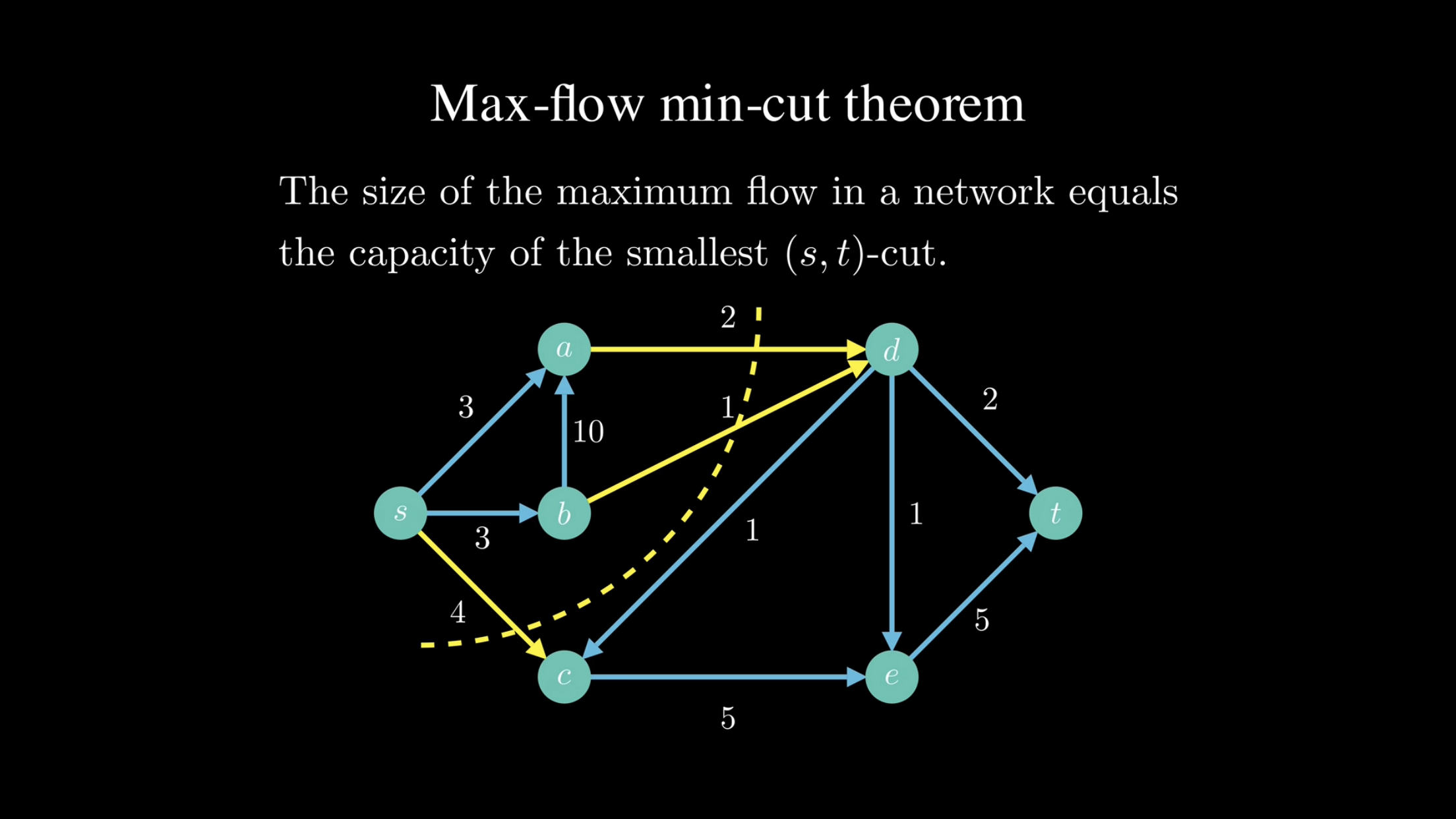}}
\caption{Max-flow min-cut theorem applied on a flow network.}
\label{fig}
\end{figure}

The feedback from viewers on YouTube, TikTok, RedNote, and Bilibili was mostly positive. Many left encouraging comments and showed support. However, criticisms can also help improve the quality of the content. Clear and concise explanations that cover main points prove to be appreciated by the audience.

\section{Manim Beyond Math and Computer Science}

So far, the examples discussed are topics from the fields of mathematics and computer science. The animation techniques can be directly extended to physics, chemistry, and other STEM subjects. In this section, examples are provided for demonstrations.

Figure 8 shows a selected frame from a magnetic field animation. The concept of a magnetic field is very important in physics, particularly in electricity and magnetism. Animating it is very straightforward with the Manim \verb|ArrowVectorField| class. Less than 50 lines of code were used. Physics is the foundation of engineering---the creation of engineering-specific animations is similar to the process of creating physics animations. 

\begin{figure}[htbp!]
\centerline{\includegraphics[width=8cm]{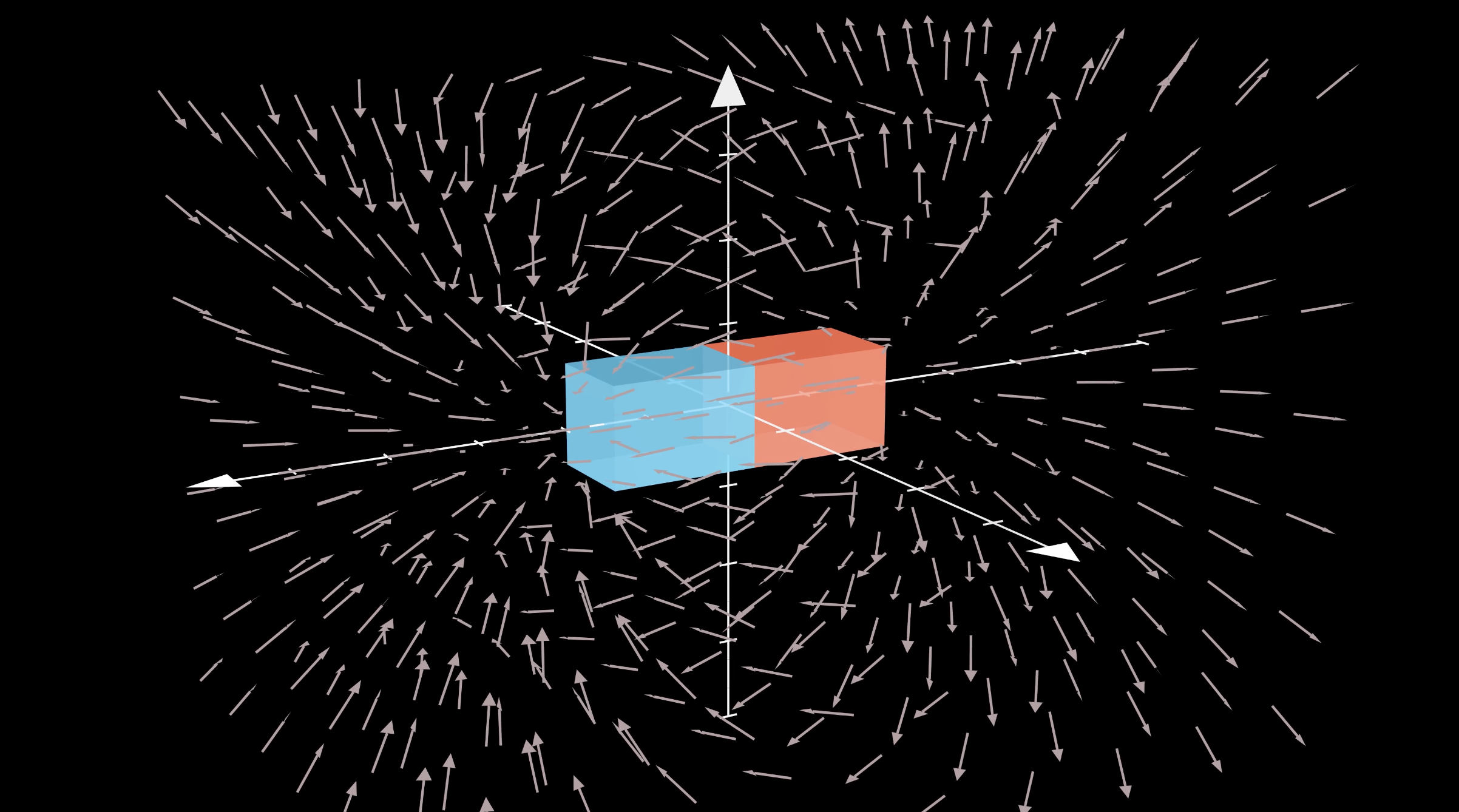}}
\caption{Magnetic field animation.}
\label{fig}
\end{figure}

Figure 9 shows the animation of a chemical reaction. The chemical equation was written using \LaTeX. The \verb|Circle| class was used to animate the molecules. The blue circles denote hydrogen, and the red circles denote oxygen. The creation of this animation only used 30 lines of code. The equation was first written, followed by the animation of the corresponding molecules. Such animations are also useful in biology, where there are a lot of chemical reactions. 

\begin{figure}[htbp!]
\centerline{\includegraphics[width=8cm]{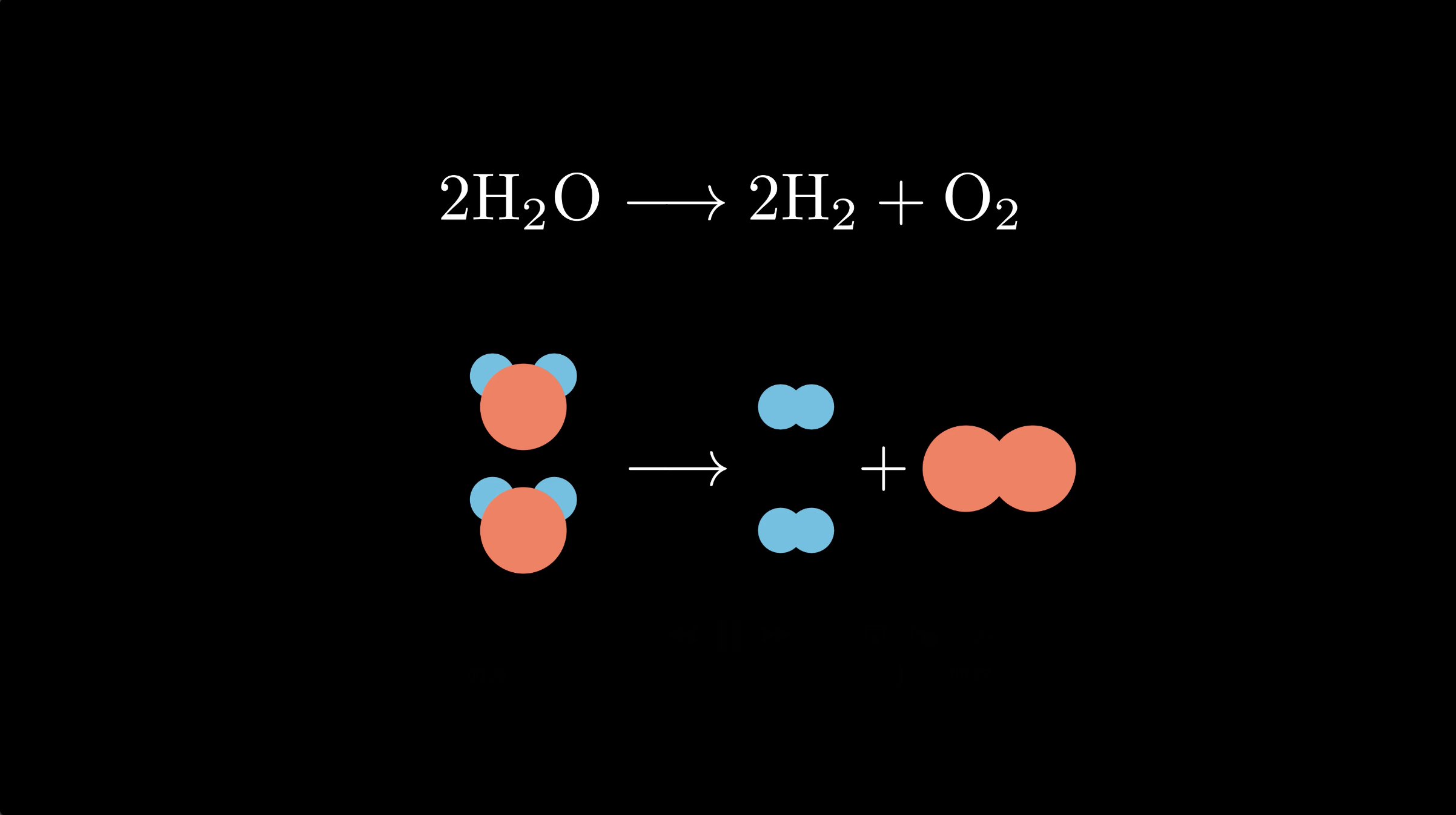}}
\caption{Chemical reaction animation.}
\label{fig}
\end{figure}

The wide variety of Manim classes and methods makes the tool very efficient in creating animations. It is clear that visualization can improve the student's understanding of various STEM topics, not limited to computer science and mathematics. The Python package Manim allows educators to generate their own visualizations, addressing the common difficulty of illustrating complex concepts by hand.



\section{Conclusion}

Manim is a very helpful tool for visualizing scientific problems through animations. The simplicity of Python's syntax and the comprehensive set of built-in Manim methods and classes enables educators to easily design and implement their own animations. Most of the feedback from viewers on YouTube, TikTok, RedNote, and BiliBili was positive. There were many encouraging comments. A few suggestions were made regarding the color and font in the visualization, offering valuable feedback for future improvements. Overall, clear and concise explanations were well-received by the audience when accompanied with engaging animations. Educators should consider adopting this visualization tool for teaching STEM courses. 

Animation frames shown in this paper are from my YouTube channel: \href{https://www.youtube.com/@ChristinaZhang-c4y}{https://www.youtube.com/@ChristinaZhang-c4y}.







\printbibliography


\end{document}